\documentstyle[12pt]{article}

 \newcommand{\be}{\begin{equation}}
 \newcommand{\ee}{\end{equation}}
 \newcommand{\ba}{\begin{eqnarray}}
 \newcommand{\ea}{\end{eqnarray}}
 
 \newcommand{\del}{\partial}

\newcommand{\lef}{\left}

\newcommand{\ri}{\right}

\newcommand{\cl}{{\cal L}}

\newcommand{\fr}{\frac}

\newcommand{\emn}{\epsilon^{\mu\nu\alpha\beta}}

\begin{document}

\begin{titlepage}

\topmargin -15mm

\rightline{\bf UFRJ-IF-FPC-008/96}

\vskip 10mm

\centerline{ \LARGE\bf Explicit Bosonization of the Massive }
\vskip 2mm
\centerline{ \LARGE\bf  Thirring Model in 3+1 Dimensions }

    \vskip 2.0cm

    \centerline{\sc R.Banerjee $^*$ and E.C.Marino }

     \vskip 0.6cm
     
\centerline{\it Instituto de F\'\i sica}
\centerline{\it Universidade Federal do Rio de Janeiro } 
\centerline{\it Cx.P. 68528, Rio de Janeiro, RJ 21945-970, Brasil} 
\vskip 2.0cm

\begin{abstract} 

We bosonize the 
Massive Thirring Model in 3+1D for small coupling constant and
arbitrary mass. The bosonized action is explicitly obtained both in
terms of a Kalb-Ramond tensor field as well as in terms of a dual
vector field. An exact bosonization formula for the current is derived.
The small and large mass limits of the bosonized theory are examined
in both the direct and dual forms. We finally obtain the exact bosonization
of the free fermion with an arbitrary mass.

\end{abstract}

\vskip 3cm
$^*$ On leave of absence from S.N.Bose National Centre for Basic
Sciences, Calcutta, India.
\vskip 3mm
Work supported in part by CNPq-Brazilian National Research Council.
     E-Mail addresses: rabin@if.ufrj.br; marino@if.ufrj.br

\end{titlepage}

\hoffset= -10mm

\leftmargin 23mm

\topmargin -8mm
\hsize 153mm
 
\baselineskip 7mm
\setcounter{page}{2}

The method of bosonization has proven to be a very powerful tool for
investigating two-dimensional theories allowing, for instance,
the obtainment of exact
solutions of nonlinear theories like Quantum Electrodynamics and Sine-
Gordon model \cite{bos}. A lot of effort
has been made in order to generalize
this method to higher dimensions \cite{rb1,bos3}
 but explicit results are only
available  in 2+1D \cite{em1,s1,rb2,kk,o1}.

In the present work we aim to provide an explicit bosonization of the
Massive Thirring Model (MTM) in 3+1D. Using the functional methods developed
in \cite{rb1} we show that the MTM for arbitrary mass and coupling can be
bosonized either in terms of a second
rank pseudotensor Kalb-Ramond  gauge field
or in terms of a dual vector gauge field. An exact bosonization formula
for the current is derived both in terms of the tensor and vector
fields. We then perform a small coupling
expansion, obtaining thereby concrete expressions for the bosonized
lagrangian and its dual for arbitrary mass. The small and large mass limits
are analyzed in detail. The leading contribution in the small mass limit
behaves as a Proca theory, either in the tensor or vector cases. In the
large mass limit, however, the leading contribution is nonlocal.
This is to be compared with the corresponding analysis in 2+1D
\cite{em1,s1,rb2,kk,o1},
where the roles of the two limits are reversed.

We also consider the case of a free fermion with an arbitrary mass as a
limiting situation of the MTM when the coupling constant vanishes. In this
case, we get exact  explicit results for the bosonized lagrangian and its
dual. Interestingly, the tensor field lagrangian
that appears in the small mass limit has been shown to be 
connected with QED \cite{em2}.  

Let us consider the master lagrangian  \cite{rb1}
\be
\cl_M = \fr{1}{6} F_{\mu\nu\alpha} F^{\mu\nu\alpha} + 
\bar\psi (i \not\del - m - \lambda \not B ) \psi +
\epsilon^{\mu\nu\alpha\beta} B_\mu\del_\nu A_{\alpha\beta}
\label{ml}
\ee
where $A_{\alpha\beta}$ is the second rank antisymmetric Kalb-Ramond
tensor field and
\be
F^{\mu\nu\alpha}= \del^{[\mu} A^{\nu\alpha]}
\label{f}
\ee
is the corresponding field intensity tensor.
$B_\mu$ is an external vector field and
$\lambda$ is its coupling constant to the fermions. This
lagrangian is invariant under independent gauge transformations on the vector
and tensor fields.
Consider the euclidean generating functional
in the presence of external sources
\be
Z= \int DB_\mu DA_{\alpha\beta} D\psi D\bar\psi
\exp \lef\{-\int d^4z[ \cl_M +
\emn \del_\nu A_{\alpha\beta} J_\mu + B_\mu K^\mu ]\ri \}
\delta [\del_\mu B^\mu] \delta [\del_\alpha
A^{\alpha\beta}]
\label{z}
\ee
where $K_\mu$ must be conserved in order to preserve gauge invariance
and the delta functions are for fixing the gauge.
Upon integration over $B_\mu$ and $A_{\alpha\beta}$ \cite{rb1}
we obtain the MTM as the resulting theory
\be
Z= \int  D\psi D\bar\psi
\exp \lef\{-\int d^4z[ 
\bar\psi (i \not\del - m ) \psi - \fr{\lambda^2}{4} j^\mu j_\mu +
\lambda j_\mu ( J^\mu + \fr{K^\mu}{2} ) ] \ri \}
\label{z1}
\ee
where $j^\mu= \bar\psi \gamma^\mu \psi$ is the fermionic current.
By comparing the generating functionals (\ref{z}) and (\ref{z1}) we can make
the following identifications
\be
\lambda j^\mu = \emn \del_\nu A_{\alpha\beta} = 2 B_\mu
\label{j}
\ee
The above results are valid for arbitrary values of $\lambda$ and $m$.
Note that the antisymmetric Kalb-Ramond field must be a pseudotensor.
This behavior under parity transformations is a general feature of the
bosonized fields in any dimension and follows from the current bosonization
formulae.

Let us perform now the fermionic integration in (\ref{z}), which yields
the familiar functional determinant. Since $B_\mu$ is an external field
only one-loop graphs will contribute to this determinant.
In the small
$\lambda$ approximation, the leading order contribution is a two-legs
graph. We therefore obtain the following effective action
\be
S_{eff} = \int d^4z \lef [ \fr{1}{6} F_{\mu\alpha\beta}F^{\mu\alpha\beta}
+ \emn B_\mu \del_\nu A_{\alpha\beta} +
\fr{1}{2} B_\mu \Pi^{\mu\nu} B_\nu
+ \emn \del_\nu A_{\alpha\beta} J_\mu \ri ]
\label{seff}
\ee
where $\Pi^{\mu\nu}$ is the lowest order contribution to the vacuum
polarization tensor of QED, which in euclidean momentum space is
given by \cite{iz}
$$
\Pi^{\mu\nu}(k) = (k^2 \delta^{\mu\nu}- k^\mu k^\nu) \Pi(k^2)
$$
where
\be
\Pi(k^2)= -\fr{\lambda_R^2}{12\pi^2 }\lef \{ \fr{1}{3} + 2
\lef (1-\fr{2m^2}{k^2}\ri ) \lef [ \fr{1}{2} x \ln \fr{x+1}{x-1} - 1 \ri]
\ri\}
\label{pi}
\ee
in which $x=\lef (1+\fr{4m^2}{k^2}\ri )^{1/2}$. In the above expression, the
renormalized coupling constant $\lambda_R$ is given, in lowest order, by
\be
\lambda_R^2 = \lef [ 1 - \fr{\lambda_R^2}{12\pi^2} \ln \Lambda^2
\ri ] \lambda^2
\label{lam}
\ee
where $\Lambda$ is an ultraviolet cutoff. Notice that in the effective action
(\ref{seff}) we have set the external source $K^\mu =0$ because after the
identification (\ref{j}), the use of two sources would be superfluous

From the effective action (\ref{seff}), we can obtain the bosonized
theory (or its dual) by integrating either over $B_\mu$ or
$A_{\alpha\beta}$. The quadractic $B_\mu$-integration can made in a
straightforward manner, giving the result
$$
Z= \int  DA_{\alpha\beta} \delta \lef[\del_\alpha A^{\alpha\beta}\ri ]
\exp \lef\{-\int d^4z \lef [ \fr{1}{6} F_{\mu\alpha\beta}F^{\mu\alpha\beta}
+\fr{1}{6} F_{\mu\alpha\beta}\lef [ G(z-z')\ri ] F^{\mu\alpha\beta}
\ri.\ri.
$$
\be
\lef.\lef.
+  \emn \del_\nu A_{\alpha\beta} J_\mu \ri ]\ri \}
\label{z2}
\ee
where the Fourier transform of $G(z-z')$ is
\be
G(k) = \fr{1}{k^2 \ \Pi(k^2)}
\label{g}
\ee
The exponent in the above integrand is the bosonized theory corresponding to
the MTM for {\it arbitrary} mass and small $\lambda$.
This is one of the central results of our paper.

Let us investigate now the small and large mass limits of the bosonized
theory. From (\ref{pi}) and (\ref{g})
 a straightforward computation yields respectively
$$
G(k) \stackrel{m \rightarrow 0} {\longrightarrow} 
\fr{36\pi^2}{5\lambda_R^2 k^2} + O\lef(\fr{m^2}{k^2}\ri)
$$
\be
G(k) \stackrel{m \rightarrow \infty} {\longrightarrow} 
- \fr{48 \pi^2 m^2}{\lambda_R^2 k^4} + O\lef(\fr{k^2}{m^2}\ri)
\label{gmm}
\ee
Observe that in the small mass limit, 
because of the gauge fixing constraint, the leading term is a
mass term for the Kalb-Ramond field. In the large mass limit, the leading
term is already nonlocal. This is a consequence of
the non-constant behavior of
the vacuum polarization tensor of QED which
must vanish for small $k$, in such a way that the
Coulomb potential has vanishingly small corrections at large distances.
Should the vacuum polarization tensor have a constant behavior for
large mass (small $k$) we would also have a Proca type lagrangian for
the bosonized theory. This may happen in higher dimensions \cite{rb1,bos3}

Let us point out that
here we have a similar situation to the three-dimensional case where
the leading behavior in
the small and large mass limits give different expressions
\cite{em1,s1,rb2,kk,o1} 
but only the role is reversed because the local form is
obtained there in the large mass case. 

Now we can get the dual version of the bosonized theory by starting from
(\ref{seff}) and performing the quadractic integration over the Kalb-
Ramond field. The result is \cite{rb1}
\be
Z= \int  DG_\mu 
\exp \lef\{-\int d^4z \lef [ \fr{1}{2} G_{\mu\nu}
\lef [\Pi(z-z')\ri] G^{\mu\nu} +
G^\mu G_\mu + 2 G_\mu J^\mu \ri ] \ri \}
\label{z3}
\ee
where $G_{\mu\nu}= \del_\mu G_\nu - \del_\nu G_\mu$
and $\Pi(z-z')$ is the inverse Fourier transform of $\Pi(k^2)$.
Observe that the first term came from the fermionic integration and was not
involved in the integration over the Kalb-Ramond field.

The small and large $m$ limits of the above expression can be obtained
trivially from the expressions in (\ref{gmm}).
Observe that in the small mass limit,
the leading contribution yields a Proca lagrangian. It is interesting to
see that in this limit both the original and dual bosonized lagrangians
are of the Proca type. The fact that the dual of a vector Proca theory
is a Kalb-Ramond Proca theory in arbitrary dimension has been observed in
\cite{rb1}.

Let us consider now the free fermion field with an arbitrary mass. This can
be obtained by taking the limit $\lambda_R \rightarrow 0$, in which case
the fermionic integration becomes exact. From (\ref{gmm}) we can see
that in this case only the second term in the exponent in (\ref{z2}) is
relevant and the bosonic action corresponding to a free massive Dirac
fermion in 3+1D is given by
\be
\int d^4z
\bar\psi (i \not\del - m ) \psi \leftrightarrow
\fr{1}{6} \int d^4z
F_{\mu\alpha\beta}\lef [\tilde G(z-z')\ri ] F^{\mu\alpha\beta}
\label{soef}
\ee
where we have rescaled the Kalb-Ramond field as $A_{\mu\nu} \rightarrow
\lambda_R^{-1}\  A_{\mu\nu}$ and therefore
\be
\tilde G(z) \equiv \lambda_R^2 G(z)
\label{tg}
\ee
It is interesting to note that the leading contribution to the
small mass limit of (\ref{soef}) reproduces the theory studied
in \cite{em2} in connection to QED,
where $G$ is proportional to $\fr{1}{\Box}$,
whereas in the large mass limit $G$
will be proportional to $\fr{1}{(\Box)^2}$.

A very important remark now is in order. Observe that according to the
identity (\ref{j}) the Thirring interaction corresponds to the first
term in (\ref{z2}). We have just seen, on the other hand, that the free
part of the fermion lagrangian can be exactly bosonized by (\ref{soef}).
One could then be tempted to exactly bosonize the full
MTM
by just adding the two pieces as it occurs in 1+1D. This
however is not true here as our computation clearly shows that only in the
$\lambda_R \rightarrow 0$ limit we can bosonize the free massive fermion
lagrangian as (\ref{soef}) and ignore the higher order insertions in the
fermionic determinant. A consequence of this
observation is that in spite of the fact
that the linear current bosonization formula (\ref{j}) is always valid,
the bosonization formula for the free fermion operator depends on the
interacting part of the theory it is embedded on and only in the free case
it is given by (\ref{soef}). This is the most remarkable point of departure
from the usual bosonization scheme in 1+1D.

We conclude by remarking that the results obtained here open a broad
field of research in the bosonization of theories in the physical dimension
of 3+1. One could devise, for instance, the use of the master lagrangian
given in \cite{rb1} in order to obtain explicit bosonization formulas for
QED. Also one could explore the concrete bosonized theories obtained here
through the operator formulation developed in 2+1D \cite{em1} for the
direct bosonization of the fermion field operator.

\vfill\eject

\leftline{\Large\bf Acknowledgements} \bigskip

Both authors  
were partially supported by CNPq-Brazilian National Research Council.
RB is very grateful to the Instituto de F\'\i sica-UFRJ 
for the kind hospitality.


\begin{thebibliography}{99}

\bibitem{bos} J.Lowenstein and J.A.Swieca, {\it Ann. of Phys.}
{\bf 68} (1971) 172;
S.Coleman {\it Phys. Rev.} {\bf D11} (1975) 2088;
S.Mandelstam, {\it Phys. Rev.} {\bf D11} (1975) 3026;
M.B.Halpern {\it Phys. Rev.} {\bf D12} (1975) 1684

\bibitem{rb1} R.Banerjee, {\it Nucl. Phys.} {\bf B465} (1996) 157

\bibitem{bos3} C.Burgees, C.L\"utken and F.Quevedo,
 {\it Phys. Lett.} {\bf B336} (1994) 18;
 P.A.Marchetti, {\it ``Bosonization and Duality in Condensed Matter
 Systems''}, hep-th/9511100 (1995);
 K.Ikegami, K.Kondo and A.Nakamura, {\it Progr. Th. Phys.} {\bf 95}
 (1996) 203

\bibitem{em1} E.C.Marino, {\it Phys. Lett. } {\bf B263 } (1991) 63

\bibitem{s1} E.Fradkin and F.Schaposnik {\it Phys. Lett. } {\bf B338 } (1994)
253;
J.C.Le Guillou, C.N\'u\~nez and F.Schaposnik, {\it ``Current Algebra and
Bosonization in Three Dimensions''} hep-th/9602017 (1996)

\bibitem{rb2} R.Banerjee, {\it Phys. Lett.} {\bf B358} (1995) 297

\bibitem{kk} K.Kondo, {\it Progr. Th. Phys.} {\bf 94} (1995) 899

\bibitem{o1} D.Barci, C.D.Fosco and L.E.Oxman {\it Phys. Lett. } {\bf B375 }
(1996)

\bibitem{em2} E.C.Marino {\it Int. J. Mod. Phys.} {\bf A9} (1994) 4009

\bibitem{iz} C.Itzykson and J.B.Zuber, {\it ``Quantum Field Theory''},
McGraw Hill, New York (1980)


                                           
\end{thebibliography}
\end{document}